\documentclass[journal,a4paper]{IEEEtran}

\usepackage{cite}
\usepackage{graphicx}
\usepackage{epstopdf}
\usepackage{amsmath}
\usepackage{amsfonts}
\usepackage{algorithmic}
\usepackage{array}
\usepackage[caption=true,font=footnotesize]{subfig}
\usepackage{stfloats}
\usepackage{multirow}
\usepackage{color}
\usepackage{amsthm}

\newtheorem{lemma}{Lemma}

\makeatletter
\g@addto@macro \normalsize {%
 \setlength\abovedisplayskip{5pt}%
 \setlength\belowdisplayskip{5pt}%
}
 \newcommand{\thickhline}{%
    \noalign {\ifnum 0=`}\fi \hrule height 1pt
    \futurelet \reserved@a \@xhline
		}
\makeatother

\begin{document}

\title{Analysis of the Decoupled Access for Downlink and Uplink in Wireless Heterogeneous Networks}

%\author{Katerina~Smiljkovikj,~\IEEEmembership{Student Member,~IEEE,}
				%Petar~Popovski,~\IEEEmembership{Senior~Member,~IEEE,}
        %and~Liljana~Gavrilovska,~\IEEEmembership{Senior~Member,~IEEE}% 

\author{Katerina~Smiljkovikj,
Petar~Popovski
and~Liljana~Gavrilovska% 				
				
\thanks{K. Smiljkovikj and L. Gavrilovska are with the Ss Cyril and Methodius University in Skopje, Macedonia (e-mail: $\left\{\mbox{katerina, liljana}\right\}$@feit.ukim.edu.mk)}%
\thanks{P. Popovski is with Aalborg University, Aalborg, Denmark (e-mail: petarp@es.aau.dk)}%
%\thanks{Digital Object Identifier}
}

\markboth{IEEE WIRELESS COMMUNICATIONS LETTERS}%
{Smiljkovikj, Popovski \MakeLowercase{\textit{and}} Gavrilovska: Decoupling of uplink and downlink transmissions}

\maketitle

\begin{abstract}
Wireless cellular networks evolve towards a heterogeneous infrastructure, featuring multiple types of Base Stations (BSs), such as Femto BSs (FBSs) and Macro BSs (MBSs). A wireless device observes multiple points (BSs) through which it can access the infrastructure and it may choose to receive the downlink (DL) traffic from one BS and send uplink (UL) traffic through another BS. Such a situation is referred to as \emph{decoupled DL/UL access}. Using the framework of stochastic geometry, we derive the association probability for DL/UL. 
In order to maximize the average received power, as the relative density of FBSs initially increases, a large fraction of devices chooses decoupled access, i.e. receive from a MBS in DL and transmit through a FBS in UL. 
We analyze the impact that this type of association has on the average throughput in the system. 
\end{abstract}

% Note that keywords are not normally used for peerreview papers.

% For peer review papers, you can put extra information on the cover
% page as needed:
% \ifCLASSOPTIONpeerreview
% \begin{center} \bfseries EDICS Category: 3-BBND \end{center}
% \fi
%
% For peerreview papers, this IEEEtran command inserts a page break and
% creates the second title. It will be ignored for other modes.
\IEEEpeerreviewmaketitle

\section{Introduction}

In the quest for better wireless connectivity and higher data rates, the cellular network is becoming heterogeneous, featuring multiple types of Base Stations (BSs) with different cell size. Heterogeneity implies that the traditional strategies in cell planning, deployment and communication should be significantly revised \cite{Andrews}.  Since the number of BSs becomes comparable to the number of devices \cite{PetarFederico5G} and the deployment pattern of the BSs is rather irregular, there are multiple BSs from which a device can select one to associate with. 

%The evolution of cellular networks is constantly progressing towards increased heterogeneity of network infrastructure. From homogeneous cellular networks consisting of macro base stations only to heterogeneous cellular networks consisting of multiple types of base stations, network engineers have plenty of challenges to deal with in order to gain the maximum fm this evolution. Heading towards fifth generation (5G) of cellular networks it is important to explore and utilize all the benefits from current cellular networks in order to be prepared for the anticipated traffic demands for 2020 era. 
%
%The exponential growth of mobile traffic shifts operators' focus from achieving coverage to achieving both, coverage and capacity. In order to meet both requirements, network operators deploy additional cells, usually small cells indoors. The motivation behind the deployment of small cells indoors lies in the fact that 2/3 of voice traffic and 90\% of data traffic occur in indoor environment \cite{ref1}. This leads to dense heterogeneous cellular networks, which meet the current coverage and capacity requirements and open new opportunities.

The key issue in a wireless heterogeneous setting is the way in which a device selects an  Access Point (AP). The authors in \cite{Andrews} and \cite{Chih-Lin} indicate that the AP selected for downlink (DL), termed Downlink AP (DLAP), is not necessarily the same as the Uplink AP (ULAP). The current cellular networks use a criterion applicable to the DL for association in both directions, i.e. a device selects the BS that offers maximal Signal-to-Interference-plus-Noise Ratio (SINR) in the DL and then uses the same BS for UL transmission. When DLAP$\neq$ULAP, we say that the device has a \emph{decoupled access}. There are two main drivers for decoupled access: (1) the difference in signal power and interference in DL as compared to UL \cite{Andrews}; and (2) the difference in congestion between BSs \cite{Chih-Lin}. Decoupled DL/UL access has been considered in \cite{Elshaer}, where the authors devise separate criteria for selection of DLAP and ULAP, respectively, and demonstrate the throughput benefits by using real-world data from planning tools of a mobile operator. Another related work that considers different associations in UL and DL is\cite{TonyD2D}, where coverage probability and throughput are analyzed for dynamic TDD networks enhanced with Device-to-Device (D2D) links.

This letter focuses on the analytical characterization of the decoupled access by using the framework of stochastic geometry \cite{Baccelli}. We use the same association criteria as in \cite{Elshaer}. We perform a joint analysis of the DL and UL association, using the same realization of the random process that describes spatial deployment of the BSs and devices. The analysis is performed for a two-tier cellular network, consisting of Macro BSs (MBSs) and Femto BSs (FBSs). This is used to obtain the central result of the paper, which is the set of association probabilities for different DL/UL configurations. The analytical results are closely matching the simulations and provide interesting insights about the decoupled access in terms of e.g. fairness regarding the UL throughput. Combining novel results from this letter with already available results in the literature, we provide an analytical justification of the phenomenon of decoupled access compared to current DL-based association in heterogeneous networks.

The letter is organized as follows. Section II describes the system model. In Section III, we derive the association probabilities and the average throughput. Section IV gives the numerical results and Section V concludes the paper. 

\section{System Model}

%The system model represents a heterogeneous cellular network consisting of two tiers of Base Stations (BSs), Macro Base Stations (MBSs) and Femto Base Stations (FBSs). 

%\subsection{Base station association}

%The devices can associate to MBS or to FBS, in both, DL and UL directions. In this analysis, DL and UL associations are analyzed separately, in terms that the decision for association in one direction does not influence the association in the other direction.
%
%Most commonly used association criteria in downlink is maximum SINR where the device associates to the BS from which it receives signal with highest SINR. This is equivalent to highest Signal-to-Noise Ratio (SNR) and highest signal power association. In this paper we assume highest average SNR association or highest average signal power association, where the averaging is performed on the fast fading process. This is valid assumption since fast variations of the fading process can lead to ping-pong effects in the association process.
%
%Uplink association adopts highest average SNR or equivalently highest average signal power in uplink. Due to the equal transmit power of all devices, uplink association is equivalent to nearest base station association. In \cite{ref4}, uplink association is termed PathLoss (PL) based association. Basically, we refer to the same association process assuming that PL model is distance based.

%\subsection{Joint system modeling}

We model a two-tier heterogeneous cellular network. The locations of BSs are modeled with independent homogeneous Poisson Point Processes (PPPs). We use $\Phi_v$ to denote the set of points obtained through a PPP with intensity $\lambda_v$, where $v=M$ for MBSs, $v=F$ for FBSs and $v=d$ for the devices. Similarly, we use $P_v$ with $v \in \{M, F, d\}$ to denote the transmission power of the node $v$. The variables $x_M, x_F \in \mathbb{R}^2$ denote the two-dimensional coordinate of MBS and FBS, respectively. The analysis is performed for a typical device located at the origin, which is the spatial point $x_d =(0,0)$.
By Slivnyak's theorem \cite{Chiu}, the distribution of a point process in $\mathbb{R}^2$ is unaffected by addition of a node at the origin. The power received by a typical device in DL from a BS located at $x_{v} \in \Phi_{v}$, where $v \in \{F,M\}$ is denoted by $\textrm{S}_{v,D}$. The power received by a BS from the typical device in UL is denoted by $\textrm{S}_{v,U}$. These powers are given by:
\begin{eqnarray}
	\textrm{S}_{v,D} = P_v h_{x_v} \left\|x_v\right\|^{-\alpha}; \textrm{S}_{v,U} = P_d h_{x_v} \left\|x_v\right\|^{-\alpha} \label{DLUL_signals}
\end{eqnarray} 
where  $\left\|x_M\right\|$ and $\left\|x_F\right\|$ are distances from the points ${x_M} \in \Phi_M$ and ${x_F} \in \Phi_F$ to the origin, respectively, and $\alpha$ is the path loss exponent ($\alpha>2$). $h_{x_v}$ is independent exponentially distributed random variable with unit mean, representing Rayleigh fading at the point $x_v$. Each receiver in the system has a constant noise power of $\sigma^2$.

The DL SINR when the device is associated to $v$BS is:
\begin{equation}
\begin{split}
	\textrm{SINR}_{v}^D = \frac{P_v h_{x_{v}} \left\|x_{v}\right\|^{-\alpha}}{\sum\limits_{{x_j} \in {\Phi}_{v} {\backslash} \{x_{v}\}} P_v h_{x_{j}} \left\|x_{j}\right\|^{-\alpha} + \sum\limits_{{x_i} \in {\Phi}_{u} } P_u h_{x_{i}} \left\|x_{i}\right\|^{-\alpha} + \sigma^2},
	\label{DL_SINR}
\end{split}	
\end{equation} 
\noindent where $v,u \in \{M,F\}$ and $v \neq u$. 
%We use both $v$ and $u$ only when there is a range of confusion. 
With the notion of typical point located at the origin, UL SINR is calculated at the location of ULAP. This involves calculation of distances between the interfering devices and ULAP, which complicates the analysis because none of them is located at the origin. The problem is solved by using the translation-invariance property of stationary point processes, by which the processes $\Phi=\{x_n\}$ and $\Phi_x=\{x_n+x\}$ have the same distribution for all $x \in \mathbb{R}^2$ \cite{Chiu}. Thus, translation of the points for the same value of $x$ preserves the process properties. We use this to shift the points for the distance between the typical device and ULAP such that the ULAP becomes located at the origin.

%\begin{figure}[!t]
%\centering
%\includegraphics[width=2.5in]{ShiftPPP}
%\caption{Original and shifted versions of PPPs.}
%\label{shiftPPP}
%\vspace{-12pt}
%\end{figure}

%The interference in UL is from devices associated to BSs different than ULAP, transmitting at the same resource as the typical device. This means that the number of interferers is equal to the number of BSs, as in \cite{NovlanULcellNet}.

The interfering devices are modeled by thinning the PPP $\Phi_{d}$ in order to take into account that only one device per BS acts as an interferer, using the same resource as the typical device \cite{NovlanULcellNet}. By thinning, we randomly select fraction of points from the original point process \cite{Chiu} with probability $p=\frac{\lambda_M+\lambda_F}{\lambda_d}$. The thinned process is denoted as $\Phi_{I_d}$ with density $\lambda_{I_d} = p \lambda_d$. The presence of a device in a Voronoi cell of a BS forbids the presence of other devices and introduces dependence among the active devices. However, this dependence is weak, as shown in \cite{NovlanULcellNet}, and it is justified to assume independent PPP for the active devices. The UL SINR at $v$BS is defined as:
\begin{equation}
\begin{split}
	\textrm{SINR}_v^{U} = \frac{P_d h_{x_{v}} \left\|x_{v}\right\|^{-\alpha}}{\sum\limits_{{x_j} \in {\Phi}_{I_d}} P_d h_{x_{j}} \left\|x_{j}\right\|^{-\alpha} + \sigma^2}
	\label{UL_SINR}
\end{split}	
\end{equation}

\section{Analysis}
\label{sec:Analysis}

The analysis is divided into two mutually related parts. We first derive the association probabilities for DL/UL and afterward use them to evaluate the average throughput.

\subsection{Association Probability}
\label{sec:AssocProb}

In DL, the device is associated to the BS from which it receives the highest average power. In UL it is associated to BS to which it transmits with the highest average power. The average power is obtained by averaging over the received signals given by~(\ref{DLUL_signals}) with respect to the fading. This is justified as the fading-induced variations can lead to ping-pong effects in the association process. The average received signal powers in DL and UL are:
\begin{eqnarray}
\mathbb{E}_h\left[\textrm{S}_{v,D}\right] = P_v \left\|x_v\right\|^{-\alpha} \label{DL_signals_ave}; \mathbb{E}_h\left[\textrm{S}_{v,U}\right] = P_d \left\|x_v\right\|^{-\alpha} \label{DLUL_signals_ave}
\end{eqnarray}
ULAP will always be the closest BS at distance $\left\|x_v\right\|$, where $v \in \{M,F\}$. The DL case is more complicated since $P_v$ is also variable. Let $x^0_v$ be the closest point to the origin from the set $\Phi_v$, with $v \in \{M,F\}$. The device is associated to 
\begin{eqnarray}
\textrm{MBS in DL if } P_M \left\|x^0_M\right\|^{-\alpha} > P_F \left\|x^0_F\right\|^{-\alpha} \label{DL_AssocRule} \\
\textrm{MBS in UL if } P_d \left\|x^0_M\right\|^{-\alpha} > P_d \left\|x^0_F\right\|^{-\alpha} \label{UL_AssocRule}
\end{eqnarray}
Otherwise, the device is associated to FBS. Let $X_v \equiv \left\|x^0_v\right\|$. The distribution of $X_v$ follows from the null probability of 2D PPP \cite{Chiu}, the probability that there is no point in the circle with radius $x$, i.e. $\Pr(X_v>x)=e^{-\pi\lambda_v x^2}$. The pdf of $X_v$ is: 
\begin{eqnarray}
	f_{X_v}(x) &=& 2\pi\lambda_v x e^{-\pi\lambda_v x^2}, x\geq0 \label{contactPDF}
\end{eqnarray}
For two-tier heterogeneous network, there are four possible combinations for choosing DLAP and ULAP:

\subsubsection{Case 1: DLAP$=$ULAP$=$MBS}
\begin{figure}[!t]
\centering
\includegraphics[width=2.4in]{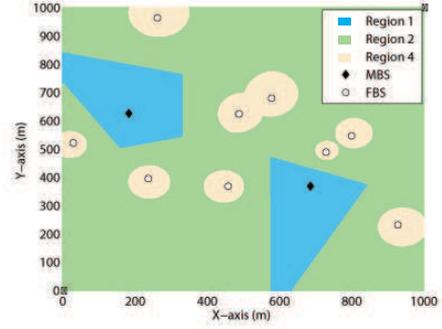}
\caption{Association regions ($P_M$=46 dBm, $P_F$=20 dBm, $\alpha$=4).}
\label{allRegions}
\vspace{-12pt}
\end{figure}

The probability that a device will be associated to MBS both in DL and UL is:
\begin{equation}
		\Pr (X_M^{-\alpha}>\frac{P_F}{P_M}X_F^{-\alpha} ; X_M^{-\alpha}>X_F^{-\alpha})
		\label{case1}
	\end{equation}
Assuming $P_F<P_M$, it follows that $P_F/P_M<1$. Therefore, the intersection of the events is the region defined by $X_M^{-\alpha}>X_F^{-\alpha}$, denoted as Region 1 on Fig. \ref{allRegions}.
The association probability of Case 1 is calculated as:
\begin{align}
		&\Pr(\textrm{Case }1) = \Pr (X_M^{-\alpha}>X_F^{-\alpha})= \nonumber \\
%							&=& P(X_F>X_M) \nonumber \\
						&= \int_0^\infty {(1-F_{X_F}(x_M))f_{X_M}(x_M)} \mathrm{d}{x_M} = \frac{\lambda_M}{\lambda_M + \lambda_F}
		\label{Pcase1}
\end{align}
The derivation of the remaining cases follows the same procedure and we thus only provide the final results.

\subsubsection{Case 2: DLAP$=$MBS and ULAP$=$FBS}

Case 2 defines decoupled access since DLAP$\neq$ULAP. The association probability is defined as:
\begin{equation}
		\Pr (X_M^{-\alpha}>\frac{P_F}{P_M}X_F^{-\alpha} ; X_M^{-\alpha} \leq X_F^{-\alpha})
	\label{case2}
\end{equation}
The domain that satisfies both events is $\frac{P_F}{P_M}X_F^{-\alpha} < X_M^{-\alpha} \leq X_F^{-\alpha}$ and is denoted as Region 2 on Fig. \ref{allRegions}. The association probability for Case 2 is equal to:	
\begin{equation}
	\begin{split}
		\Pr (\textrm{Case }2) = \frac{{\lambda_F}}{{\lambda_F} + {\lambda_M}} - \frac{{\lambda_F}}{{\lambda_F} +  \left(\frac{P_M}{P_F}\right)^{2/\alpha} {\lambda_M}}
		\label{Pcase2}
  \end{split}
	\end{equation}

\subsubsection{Case 3: DLAP$=$FBS and ULAP$=$MBS}

The association probability for Case 3 should satisfy the following conditions:
	\begin{equation}
		X_M^{-\alpha} \leq \frac{P_F}{P_M}X_F^{-\alpha} \cap X_M^{-\alpha} > X_F^{-\alpha}
		\label{case3}
	\end{equation}
The intersection (\ref{case3}) is an empty set and therefore the probability of DLAP$=$FBS and ULAP$=$MBS is $\Pr (\textrm{Case }3)=0$.

\subsubsection{Case 4: DLAP$=$ULAP$=$FBS}

The probability for associating to FBS in both DL and UL is defined as: 
\begin{equation}
		\Pr (X_F^{-\alpha} \geq \frac{P_M}{P_F}X_M^{-\alpha};X_F^{-\alpha} \geq X_M^{-\alpha})
		\label{case4}
\end{equation}
Since $P_M/P_F>1$, the intersection of the events is $X_F^{-\alpha} \geq \frac{P_M}{P_F}X_M^{-\alpha}$, denoted as Region 4 on Fig. \ref{allRegions}. The association probability for Case 4 is equal to:
\begin{equation}
		\Pr (\textrm{Case }4) = \frac{\lambda_F}{\lambda_F + \left(\frac{P_M}{P_F}\right)^{2/\alpha} \lambda_M}
		\label{Pcase4}
\end{equation}

\subsection{Average throughput}

The average throughput for devices associated to $v$BS with $v \in \{M,F\}$ in $m$L direction using $n$L association rules, with $m,n=D$ for DL and $m,n=U$ for UL, is calculated as:
\begin{equation}
\textrm{R}_{v,m,n}(\gamma_{th}) = \frac{1}{N_{v,n}}\textrm{log}_2(1+\gamma_{th})\textrm{P}_{c,v,m}(\gamma_{th}) %
\label{ThroughputDef}
\end{equation}
where $\gamma_{th}$ is target SINR, $\textrm{P}_{c,v,m}$ is the probability that the instantaneous SINR is greater than $\gamma_{th}$ and $N_{v,n}$ is the average number of associated devices on $v$BS and is equal to $N_{v,n}=\lambda_d A_{v,n}/\lambda_v$, with $A_{v,n}$ being the association probability for $v$BS using $n$L association rules. Using the association probabilities derived in Section \ref{sec:AssocProb}, $A_{v,n}$ is expressed as:
\begin{eqnarray}
A_{M,D} &=& \Pr(\textrm{Case }1)+\Pr(\textrm{Case }2) \nonumber \\
A_{F,D} &=& \Pr(\textrm{Case }3)+\Pr(\textrm{Case }4) \nonumber \\
A_{M,U} &=& \Pr(\textrm{Case }1)+\Pr(\textrm{Case }3) \nonumber \\
A_{F,U} &=& \Pr(\textrm{Case }2)+\Pr(\textrm{Case }4)
\end{eqnarray} 

In order to calculate the average throughput in a two-tier network, we first need to calculate the distribution of the distance to the serving BS, which depends on the association process. For DL association rules, given by (\ref{DL_AssocRule}), the pdf of the distance to the serving BS is derived in \cite{AndrewsHetNet} and is given by:
\begin{eqnarray}
f_{X_{v,D}}(x) &=& \frac{2\pi\lambda_v}{A_{v,D}}x e^{-\left( \lambda_v+\lambda_u \left( \frac{P_u}{P_v}\right)^{2/\alpha}  \right) \pi x^2 }
\end{eqnarray}
For UL association rules, given by (\ref{UL_AssocRule}), the pdf of the distance to the serving BS is given by:
\begin{eqnarray}
f_{X_{v,U}}(x) &=& \frac{2\pi\lambda_v}{A_{v,U}}x e^{-\left( \lambda_v+\lambda_u \right) \pi x^2 }
\end{eqnarray}
  
The average throughput in the DL is calculated as:
\begin{eqnarray}
\textrm{R}_{D}(\gamma_{th}) = \textrm{R}_{M,D,D}(\gamma_{th})A_{M,D} + \textrm{R}_{F,D,D}(\gamma_{th})A_{F,D}
\end{eqnarray} 
where $\textrm{R}_{M,D,D}(\gamma_{th})$ and $\textrm{R}_{F,D,D}(\gamma_{th})$ are expressed by the general formula given by (\ref{ThroughputDef}). Using the approach derived in \cite{AndrewsHetNet}, we derive the final expression for the average throughput in DL on $v$BS:
\begin{eqnarray}
\textrm{R}_{v,D,D}(\gamma_{th}) = \frac{1}{N_{v,D}}\textrm{log}_2(1+\gamma_{th}) \frac{2\pi\lambda_v}{A_{v,D}}x \times \nonumber \\ 
\int\limits_{0}^\infty e^{-\frac{\gamma_{th}\sigma^2x^\alpha}{P_v}} {e^{- \left( \lambda_v+\lambda_u \left( \frac{P_u}{P_v}\right)^{2/\alpha} \right) \left( 1+\kappa(\alpha,\gamma_{th})\right) \pi x^2}} \mathrm{d}{x}
\end{eqnarray}
where $\kappa(\alpha,\gamma_{th}) = \gamma_{th}^{2/\alpha} \int\limits_{\gamma_{th}^{-2/\alpha}}^\infty {\frac{1}{1+u^{\alpha/2}}} \mathrm{d}{u}$.
The key point in the evaluation is the following observation: if the device is associated to MBS located at $x$ the interfering MBSs are at a distance greater than $x$ and the interfering FBSs are at a distance greater than $\left( {P_F}/{P_M} \right)^{1/\alpha}x$; if the device is associated to FBS located at $x$, the interfering FBSs are at a distance greater that $x$ and the interfering MBSs are at a distance greater than $\left( {P_M}/{P_F} \right)^{1/\alpha}x$.

The average downlink throughput can be also calculated in a more elegant way by using the following:
\vspace{-6pt}
\begin{lemma} (Equivalent downlink model) \emph{A two-tier heterogeneous model is equivalent to a novel homogeneous model with BSs deployed by PPP $\widetilde{\Phi}_{MF}$ with intensity $\widetilde{\lambda}_{MF} = (\lambda_{M}+\lambda_F) \left(P_F^{2/\alpha} \frac{\lambda_F}{\lambda_M + \lambda_F} + P_M^{2/\alpha} \frac{\lambda_M}{\lambda_M + \lambda_F}\right)$. Then, the average throughput in DL can be calculated as:}
\begin{equation}
\begin{split}
	\widetilde{\textrm{R} }_{D}(\gamma_{th}) = \frac{1}{\widetilde{N}_{MF}}\log_2(1+\gamma_{th}) \int\limits_{0}^{\infty} {e^{-\gamma_{th} \sigma^2 x^\alpha}} \times \\
{2\pi \widetilde{\lambda}_{MF} x e^{-(\kappa(\gamma_{th},\alpha) + 1)\pi \widetilde{\lambda}_{MF} x^2}} \mathrm{d}{x}%
	\label{DL_Pc_final}  
\end{split}	
\end{equation}
where $\widetilde{N}_{MF}=\lambda_d/\widetilde{\lambda}_{MF}$.
\end{lemma}
\begin{IEEEproof} The DL signal power at the typical point can be represented as $\textrm{S}_{D}=Z h \left\| x_{MF}\right\|^{-\alpha}=h \left\|Z^{-\frac{1}{\alpha}} x_{MF}\right\|^{-\alpha}=h \left\|y_{MF}\right\|^{-\alpha}$, where Z is a discrete random variable with two possible values, $P_M$ and $P_F$, with probabilities $\Pr(P_M)=\lambda_M/(\lambda_M+\lambda_F)$ and $\Pr(P_F)=\lambda_F/(\lambda_M+\lambda_F)$, respectively. The points $x_{MF}$ are from a PPP $\Phi_{MF}$ with density $\lambda_{MF}=\lambda_{M}+\lambda_{F}$. By equivalence theorem \cite{Błaszczyszyn}, the spatial points $y_{MF}$ form new PPP with density $\widetilde{\lambda}_{MF} = \lambda_{MF}\mathbb{E}\left[Z^{2/\alpha}\right]$. 
\end{IEEEproof}

The average throughput in UL without decoupled access is calculated using DL association rules given by (\ref{DL_AssocRule}):
\begin{eqnarray}
\textrm{R}_{U}(\gamma_{th}) = \textrm{R}_{M,U,D}(\gamma_{th})A_{M,D} + \textrm{R}_{F,U,D}(\gamma_{th})A_{F,D}
\end{eqnarray}
\noindent where $\textrm{R}_{M,U,D}(\gamma_{th})$ and $\textrm{R}_{F,U,D}(\gamma_{th})$ are evaluated as:
\begin{eqnarray}
\textrm{R}_{v,U,D}(\gamma_{th}) = \frac{1}{N_{v,D}}\textrm{log}_2(1+\gamma_{th}) \int\limits_{0}^\infty {e^{-\frac{\gamma_{th}\sigma^2x^\alpha}{P_d}} } \times \nonumber \\
{e^{- \pi \lambda_{I_d} \kappa(\alpha,\gamma_{th}) x^2 }} f_{X_{v,D}}(x) \mathrm{d}{x}
\end{eqnarray}
It can be observed that the distribution to the serving BS and the association probabilities are from DL association rules.

The average throughput in UL with decoupled access is:
\begin{eqnarray}
\textrm{R}_{U}^d(\gamma_{th}) = \textrm{R}_{M,U,U}^d(\gamma_{th})A_{M,U} + \textrm{R}_{F,U,U}^d(\gamma_{th})A_{F,U}
\end{eqnarray}

\noindent where $\textrm{R}_{M,U,U}^d(\gamma_{th})$ and $\textrm{R}_{F,U,U}^d(\gamma_{th})$ are evaluated as:
\begin{eqnarray}
\textrm{R}_{c,v,U}^{d}(\gamma_{th}) = \frac{1}{N_{v,U}}\textrm{log}_2(1+\gamma_{th}) \int\limits_{0}^\infty {e^{-\frac{\gamma_{th}\sigma^2x^\alpha}{P_d}} } \times \nonumber \\
e^{- \pi \lambda_{I_d} \kappa(\alpha,\gamma_{th}) x^2 } f_{X_{v,U}}(x) \mathrm{d}{x}
\end{eqnarray}

\textbf{Remark 1} (Equivalent uplink model) \emph{A two-tier HetNet model with homogeneous devices is represented by an equivalent homogeneous model with BSs deployed by PPP ${\Phi}_{MF}$ with intensity ${\lambda}_{MF} = {\lambda_M + \lambda_F}$. This is a consequence of the UL association rule, which is based on path-loss only. Then, the average throughput in UL can be elegantly calculated as:}
\begin{eqnarray}
	\widetilde{\textrm{R}}_{U}^d = \frac{1}{{N}_{MF}}\textrm{log}_2(1+\gamma_{th})\int\limits_{0}^{\infty} e^{- \frac{\gamma_{th}\sigma^2x^\alpha}{P_d} } \times \nonumber \\
	{2\pi {\lambda}_{MF} x e^{-( {\lambda}_{MF} + {\lambda}_{I_d}\kappa(\gamma_{th},\alpha))\pi x^2}}  \mathrm{d}{x}%
	\label{UL_Pc_final}  	
\end{eqnarray}

The throughput gain for Case $i$, $i\in\{1,2,3,4\}$, is defined as the ratio between the throughput achieved with and without decoupling and is denoted as $\eta(\textrm{Case }i)$. The average throughput gain is calculated as: $\overline{\eta} = \sum \limits_{i=1}^4 \Pr(\textrm{Case }i) \eta(\textrm{Case }i)$.

\section{Numerical Results}

\begin{figure}[!t]
\centering
\includegraphics[width=8cm]{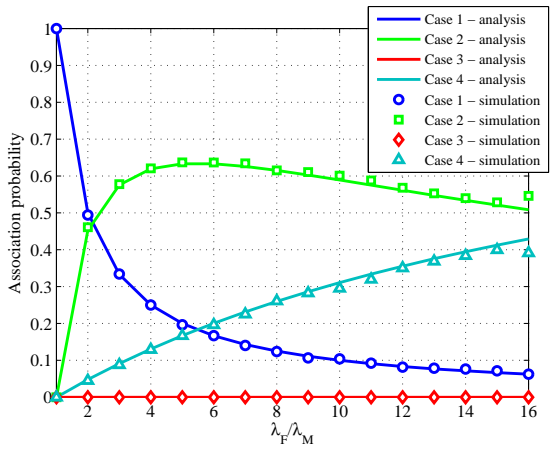}
\caption{Joint association probabilities for Cases 1-4 ($P_M$=46 dBm; $P_F$=20 dBm; $\alpha$=4).}
\label{JAP}
\vspace{-12pt}
\end{figure}

The association probabilities for each of the cases are equal to the percentage of devices that will be associated with the particular case. Fig.~\ref{JAP} shows the association probabilities for different densities of FBSs and it gives an important information about DL/UL decoupling. The percentage of devices that choose decoupled access of Case 2 (DL through MBS and UL through FBS) increases rapidly by increasing the density of FBSs. As the density of FBSs increases further, the probability for decoupled access starts to decrease slowly at the expense of increased probability for Case 4. There is a region of interest for $\lambda_F$ with a high percentage $>60\%$ of devices for which the decoupled access is optimal. As $\lambda_F / \lambda_M \rightarrow \infty$, the probability of decoupled access will go to zero.  

%\begin{figure}[!t]
%\centering
%\includegraphics[width=8cm]{CoverageProbability_PL4_new}
%\caption{Coverage probability in DL and UL $(P_M=46dBm; P_F=20dBm; P_d=20dBm; \alpha=4; \lambda_F=6\lambda_M; \sigma^2=N_0W; N_0=-174dBm/Hz; W=180kHz)$.}
%\label{CP}
%\vspace{-12pt}
%\end{figure}

%Fig.~\ref{fig:3b} shows the results for coverage probability. It is clearly visible that UL coverage probability with decoupling is significantly higher compared to the case without decoupling, with gain up to 30\%. The simulation with ``exact deployment'' refers to an accurate simulation of the UL interference, consisting of deploying $N_d$ devices, associating each of them and for each (M/F)BS, select randomly one of the associated devices to be a transmitter, causing interference to the typical device. The results show that the modeling the interference with thinning, closely approximates this accurate simulation. 

\begin{figure}[!t]
\centering
\includegraphics[width=8cm]{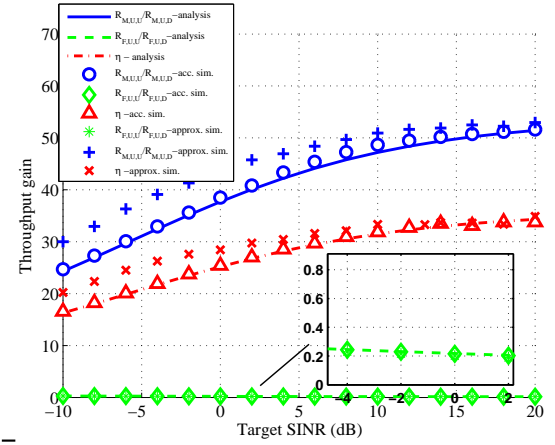}
\caption{UL throughput gain analysis, compared with an accurate simulation of devices (acc. sim.) and approximation by independent active devices (approx. sim.) ($P_M$=46dBm; $P_F$=20dBm; $P_d$=20dBm; $\alpha$=4; $\lambda_F$=10$\lambda_M$; $\sigma^2$=$10^{-12}$).}
\label{ULthroughput}
\vspace{-12pt}
\end{figure}

Fig.~\ref{ULthroughput} shows the throughput gain for the devices associated to (M/F)BSs and the average gain. There is a difference between, on one side, the accurate simulation of the devices with PPP $\Phi_{d}$, and, on the other side, its approximation by simulation of independent PPP $\Phi_{I_d}$ for the active devices only. While the UL coverage probability with DL/UL decoupling is strictly superior, the congestion of the BSs affects the throughput in a different manner. Basically, FBSs have significantly small DL coverage and therefore associate very small number of devices compared to MBSs, but each device gets higher throughput. It is visible that for $\gamma_{th}$=2 dB the throughput achieved on MBSs is 40 times higher with decoupling, while the throughput achieved on FBSs is 5 times lower with decoupling. The average throughput gain is always positive. It can be concluded that by DL/UL decoupling the devices with low SINR (located in Regions 1 and 2) achieve significant improvement, at the expense of marginal decrease in the UL throughput of the devices in Region 4. This suggests that decoupled access can be used as a tool towards achieving fairness among the accessing devices. 

\section{Conclusion}

This letter considers the problem of device association in a heterogeneous wireless environment. The analysis is done using models based on stochastic geometry. The main result is that, as the density of the Femto BSs (FBSs) increases compared to the density of the Macro BSs (MBSs), a large fraction of devices chooses to receive from a MBS in the downlink (DL) and transmit to a FBS in the uplink (UL). This is the concept of \emph{decoupled access} and challenges the common approach in which both DL and UL transmission are associated to the same BS. It is shown that the decoupling of DL and UL can be used as a tool to improve the fairness in the UL throughput. Part of our future work refers to the architecture for decoupled access, which includes signaling and radio access protocols.

% Can use something like this to put references on a page
% by themselves when using endfloat and the captionsoff option.
\ifCLASSOPTIONcaptionsoff
  \newpage
\fi

\end{document}